# SFCW GPR TREE ROOTS DETECTION ENHANCEMENT BY TIME–FREQUENCY ANALYSIS IN TROPICAL AREAS

*Wenhao Luo[1], Yee Hui Lee[1], Abdulkadir C. Yucel[1], Genevieve Ow[2] and Mohamed Lokman Mohd Yusof[2]*
[1]School of Electrical & Electronic Engineering, Nanyang Technological University, Singapore
[2]Centre for Urban Greenery & Ecology, National Parks Board, Singapore

*Abstract*— Accurate monitoring of tree roots using ground penetrating radar (GPR) is very useful in assessing the trees' health. In high moisture tropical areas such as Singapore, tree fall due to root rot can cause loss of lives and properties. The tropical complex soil characteristics due to the high moisture content tends to affect penetration depth of the signal. This limits the depth range of the GPR. Typically, a wide band signal is used to increase the penetration depth and to improve the resolution of the GPR. However, this broad band frequency tends to be noisy and selective frequency filtering is required for noise reduction. Therefore, in this paper, we adapt the stepped frequency continuous wave (SFCW) GPR and propose the use of a Joint time frequency analysis (JTFA) method called short-time Fourier transform (STFT), to reduce noise and enhance tree root detection. The proposed methodology is illustrated and tested with controlled experiments and real tree roots testing. The results show promising prospects of the method for tree roots detection in tropical areas.

## 1. INTRODUCTION

Ground Penetrating Radar (GPR), an effective non-destructive, portable testing (NDT) method, has been utilized in various areas of civil and environmental engineering [1]. A step frequency continuous waveform (SFCW) ground penetrating radar (GPR), which produces a very wideband signal that covers both the low frequency region and the high frequency region at the same time is used. In tropical areas, due to the high moisture content of the soil, one major challenge is: using the wide frequency span SFCW GPR for larger depths will not be the optimum solution. The high frequencies tends to result in noise at larger depths while low frequency mask out details at smaller depths [2]. Thus, a full bandwidth processing will not produce satisfactory radargrams over the whole depth detection range.

The joint time-frequency analysis (JTFA) [3], which can help obtain both time and frequency information from radar images has been studied for radar detection. JTFA provides various transformation algorithms between time-domain signal and frequency-domain signal, such as Gabor transform, short-time Fourier transform (STFT) and wavelet transform [3]. Of these transformation algorithms, STFT are used to locate buried targets and characterize the properties of the targets in the time-frequency domain [4], which produces a marginal resolution but at a relatively lower computational cost, which is suitable for a real-time underground tree root monitoring. In [5], they have demonstrated that by extracting the peak frequency in the spectrograms at a particular time/depth with STFT and generating a slice view of these peak frequencies, the location of the underground plastic pipe network can be determined. Recently, some studies worked on adopting STFT to assess and monitor street trees in a rapid and non-invasive way. In [6] researchers talked about the advances in the use of the STFT for assessing urban trees' root systems. In [7], a frequency spectrum-based processing framework was proposed to access the tree roots.

In this paper, a method based on the joint time-frequency analysis of the experimental data is introduced. The aim is to enhance the GPR detection capability of the tree roots in tropical areas, especially the roots buried in deep sub-surfaces. The logic of the method is illustrated with some controlled experiment in a sandpit, and a possible application of the method is verified with a set of real root testing data in the tropical area. The contribution of the paper is: 1. proposing a frequency response distribution of the B-scan data by using STFT analysis of each A-scan. 2. According to the frequency response distribution, we select the suitable frequency band of the SFCW UWB GPR to get a better B-scan in deeper sub terrains.

## 2. METHODOLOGY

### 2.1. Test Field and Equipment

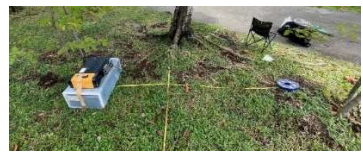

Fig. 1 The GPR survey conducted near a real tree in Singapore.

To illustrate the methodology that we proposed in a clear way, some sets of controlled comparative experiments were carried out in a sandy field, where three metal bars were buried in different depth and a tree root was buried. Meanwhile, a real tree testing survey was performed within the premises of the National Parks Board, Singapore (Figure 1). Five linear scans, each 2.3 m long and spaced 0.20 m apart from one another, were performed on the soil field near the tree trunk.

The GPR system contains a Keysight P5021A Vector Network Analyzer (VNA) (working range 9 kHz-6.5 GHz) as a transceiver and two compact dual-polarized Vivaldi antennas [8], the antennas are sealed in a box, as shown in Fig. 1. The two antennas have a 0.1 m separation. Absorbers are placed around the antennas to reduce direct coupling between the antennas and to reduce environmental noise. The VNA sweeps 1001 frequency points in the frequency band from 0.2 GHz to 4.0 GHz to record as much information as possible in the permitted working range of the VNA and antennas. The intermediate frequency bandwidth (IFBW) is set as 1000 Hz [9], and the power is set as -10 dBm. In the measurement, the transmission coefficient is recorded with a moving step of 0.03m, discretized across 71 samples, and then transformed to the time domain by IFFTs.

**2.2. Data Processing**

*2.2.1. Basic processing*

The basic processing of the collected data includes the Zero-offset removal, time-zero correction, background removal with Singular Value Decomposition (SVD), the purpose is to reduce the background noise and enrich the reflection information of the subterranean roots. Details of the basic processing can be found in [10].

*2.2.2. Short-Time Fourier Transform*

After the basic processing of the time domain data, STFT analysis was conducted to verify the frequency response of the underground objects. STFT is a method to transform time domain signals to time-frequency domain to describe how the reflected signal's frequency information changes over time. The STFT has demonstrated that the two-dimensional time-frequency analysis is effective in characterizing materials [4]. Time domain data are divided into small time windows. Each windowed data are Fourier transformed and stacked together to obtain a 2D time-frequency plot. The discrete STFT is formulated as [3]:

$$X(\tau, \Omega) = \int_{-\infty}^{\infty} x(\tau) w(\tau - t) e^{-j\Omega\tau} d\tau$$

where x is the received GPR signal, $\Omega$ is the radial frequency whose resolution $(\Delta\Omega = 2\pi/N)$ is determined by the number of points ($N$) adopted for FFT computation. In this analysis, $N$ = 1001. $\tau$ is the time resolution. Because our GPR digitizer's sampling frequency is 3.8 Gsps, $\tau$ equals 263 ps. $w(t)$ is the window function. Here a Hamming window is employed. In STFT analysis, there exists a trade-off between time and frequency resolution when determining the window size. Through a series of iterative experiments, we select 1/10 the total number of time index to set the window size. This window size gives a good balance between frequency and time resolution for the detection of cylindrical objections with radius between 2cm to 10cm.

*2.2.3. Frequency response distribution*

In each A-scan STFT result matrix, maximal value of each row (time axis) is extracted, and the extracted values form a vector, through which, the strongest reflected signals along the time axis are identified.

To minimize the surrounding noise, a threshold (a quarter of the maximum B-scan STFT data, spectrum of which is not distinguishable in time-frequency heatmap) was then defined. The STFT data below the threshold in each A-scan were set to zero. After mapping the nonzero data to their corresponding frequency points and combining all these frequency vectors of every A-scans, a 2D datasets indicates frequency response distribution is obtained.

The above process shows us how the buried objects response to the wide band SFCW signal.

**3. RESULTS AND DISCUSSION**

**3.1. Time-Frequency Domain Characteristics of Different Objects Buried in the Sandy Field**

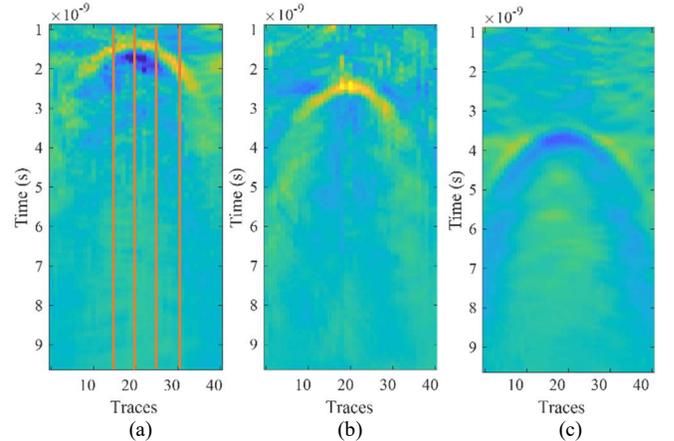

Fig. 2. Images after the basic processing. Targets buried at the depth of (a) 10cm, (b) 20cm, (c) 30cm.

In the experiments, we firstly conducted B-scans of a metal cylinder buried in the sand. The relative permittivity of the sand ($\varepsilon_{sand}$) is about 4 by averaging the values from 10 different points on the test site using N-1501A dielectric probe kit. The depth of the buried object is varied from 10 cm to 20 cm to 30 cm. The three B-scan collected and processed are shown in Fig. 2. The basic signal processing of the GPR B-scan data were carried out with 41 data points and 10 ns time window set. In Fig.2, the hyperbolas of the cylinder buried in the sand pit at different depths can be seen clearly.

STFT method was then applied on each A-scan data. Extracted from B-scan image of Fig. 2(a), Fig. 3 show the STFT results of the A-scan data that corresponds to the 15th, 20th, 25th and 30th traces in Fig. 2(a). In Fig. 3(a), a strong maximum region occurs around 2.4 ns on the 15th trace when the A-scan is performed on the left side of the buried cylinder. Fig. 3(b) and (c) show that when the A-scans are almost right above the buried cylinder (the 20th and 25th trace), strong maximum regions are produced around 1.4 ns and 1.6ns respectively. Fig. 3(d) shows a strong maximum region around 1.8 ns on the 30th trace when the A-scan is on the right side of the buried cylinder. The time of the strongest peak of each A-scan agrees well with the time of the received return signal from the buried cylinder. In Fig.3, the spectrum of the rebar is obviously seen in the frequency range of 0.4-2.5 GHz, which means strong reflections are obtained from the buried cylinder within this frequency range. Conversely, if there is a strong resonance in the time-frequency domain, there should be reflection of the target at the corresponding time point in the time domain B-scan.

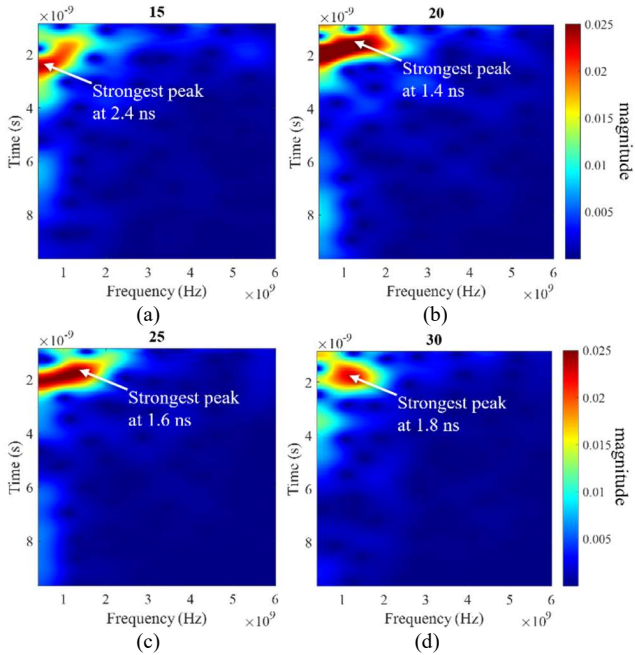

Fig. 3. STFT results of the selected A-scan data (a) trace of 15$^{th}$; (b) trace of 20$^{th}$; (c) trace of 25$^{th}$; (d) trace of 30$^{th}$.

The processed frequency response distribution results of the three B-scans in Fig. 2 are plotted in Figs. 4. It is obvious that the frequency response agrees with the hyperbola in the time domain of the corresponding B-scans. The frequency response of the targets is in the range of 0.4GHz to 2.5 GHz.

The comparison between the time-domain B-scan and frequency response distribution of the roots buried in the sandpit is shown in Fig.5. The hyperbola is no as clear as the metal cylinder but that is as expected. However, similar conclusions can be drawn. The frequency response of the target is in the range of 0.4GHz to 2.5 GHz.

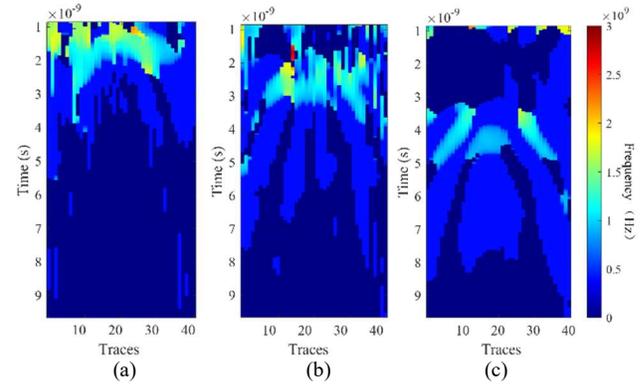

Fig. 4. Frequency response distribution of the B-scan data. Targets buried at the depth of (a) 10cm, (b) 20cm, (c) 30cm.

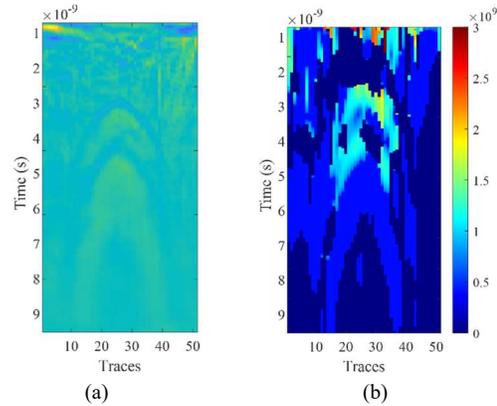

Fig. 5. Image of a root buried in the sandpit. (a) B-scan after basic process; (b) Frequency response distribution of the B-scan data.

### 3.2. The Possible Application of the Method in Enhancing the Real Tree Roots Reflection Patterns

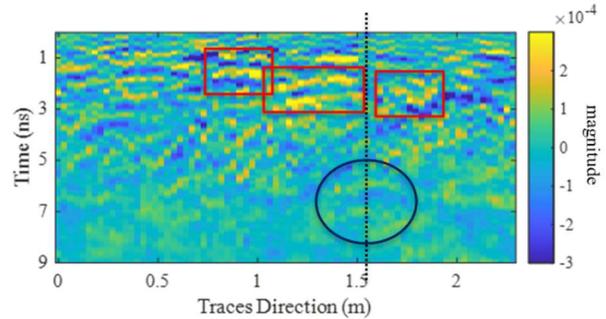

Fig. 6. B-scan after basic process of real tree root testing.

Figure 6 shows one of the B-scans of the real tree testing after the application of basic processing. Where $\varepsilon_{soil}$ is about 13. From Fig.6, some hyperbolae indicating various tree roots above 4 ns can be seen, labelled in red frame. Due to the unevenness of the soil surface, it is reasonable that some of the hyperbolae are of irregular shape. The STFT function was applied to each A-scan.

Trace of x = 1.6 m is selected for discussion. In Fig. 6, we cannot see any significant reflection in this A-scan, and no clear hyperbola can be seen below 4 ns around this A-scan.

The STFT analysis result of the A-scan is shown in Fig. 7, an obvious resonance is shown in red frame between time 5 ns and 7 ns. This indicates that there should be a reflection from roots. The reason why the reflection signal in time domain is not clear could be that the high-frequency signals cannot penetrate deep enough, the reflected signals within these frequency bands are within the noise level, which would disturb the signals from lower bands.

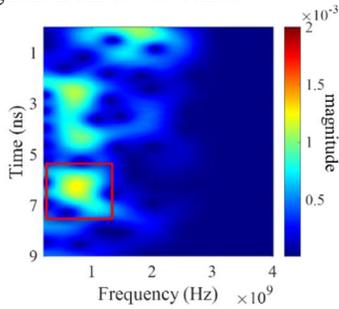

Fig. 7. STFT results of the selected A-scan trace of x = 1.6 m.

The processed frequency response distribution results of the B-scan in Fig. 6 are plotted in Fig. 8. The frequency response above 4 ns indicated that there are a number of resonances, as seen in Fig. 6. There are some significant resonances below 4ns, such as the area in the red frame around the A-scan trace of x = 1.6m, but no obvious features in black circled area in Fig.6. The frequency response distribution of the targets is in the range of 0.4 GHz to 2.0 GHz.

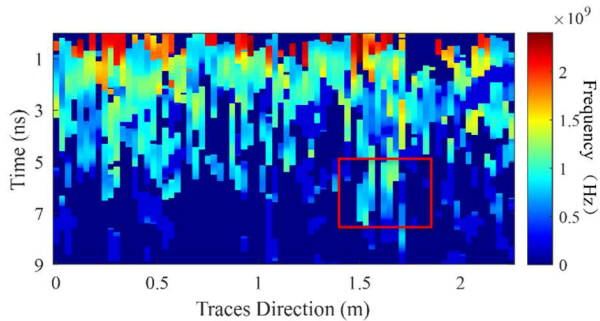

Fig. 8. Frequency response distribution of the B-scan data.

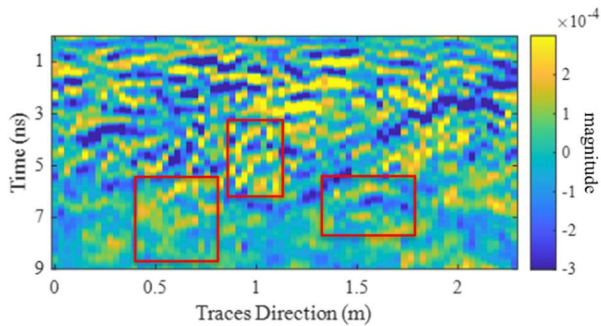

Fig. 9. Processed B-scan with selected frequency band.

By filtering out the frequencies other than the frequency response distribution in the full band, and getting the processed B-scan, we can easily see the reflection patterns from roots later than 4ns, as indicated in the red frames in Fig.9.

## 4. CONCLUSIONS AND FUTURE DEVELOPMENTS

In this paper, an advanced approach for better detection of tree roots in tropical soil environment using SFCW Ground Penetrating Radar (GPR) technology is presented. A Short-Time Fourier Transform (STFT) used to present characteristics of reflected data in both time and frequency domain, was applied to GPR data collected in controlled sand pit experiment. A frequency response distribution of the B-scan was obtained by performing STFT on all A-scan. This enables us to determine the depth of the buried targets. The proposed approach was then applied to a real tree testing data in the tropical region. Results demonstrate the effectiveness of the method in enhancing reflection patterns of tree roots. This is promising for future study of deeply buried tree roots like the permittivity and radius estimation.